# MAXIMUM F1-SCORE TRAINING FOR END-TO-END MISPRONUNCIATION DETECTION AND DIAGNOSIS OF L2 ENGLISH SPEECH

*Bi-Cheng Yan, Hsin-Wei Wang, Shao-Wei Fan Jiang, Fu-An Chao, Berlin Chen*

National Taiwan Normal University, Taipei,Taiwan
{bicheng, hsinweiwang, swfanjiang, fuann, berlin }@ntnu.edu.tw

## ABSTRACT

End-to-end (E2E) neural models are increasingly attracting attention as a promising modeling approach for mispronunciation detection and diagnosis (MDD). Typically, these models are trained by optimizing a cross-entropy criterion, which corresponds to improving the log-likelihood of the training data. However, there is a discrepancy between the objectives of model training and the MDD evaluation, since the performance of an MDD model is commonly evaluated in terms of F1-score instead of phone or word error rate (PER/WER). In view of this, we in this paper explore the use of a discriminative objective function for training E2E MDD models, which aims to maximize the expected F1-score directly. A series of experiments conducted on the L2-ARCTIC dataset show that our proposed method can yield considerable performance improvements in relation to some state-of-the-art E2E MDD approaches and the celebrated GOP method.

## 1. INTRODUCTION

Computer-assisted pronunciation training (CAPT) systems have become increasingly popular and are used for an ever-expanding set of use cases [1, 2]. In common CAPT systems, L2 (second language) learners are presented with a text prompt and asked to read it aloud. According to the input speech, CAPT systems can inform learners about mispronounced phone segments, and in turn learners can repeat the reading-aloud training so as to improve their pronunciation skills. As the core of CAPT systems, mispronunciation detection and diagnosis (MDD) module can be used to detect pronunciation errors in an L2 learner's articulation and further provide specific diagnosis feedback. The MDD module is expected to not only evaluate pronunciation proficiency at the supra-segmental level [3, 4] (e.g., fluency and intonation), but also mispronunciations at the segmental level [5, 6] (e.g., phone and word) could also be detected and diagnosed. Due to the inherent difficulty of the former, the majority of past work has focused on the latter.

There have been considerable research endeavors conducted to the detection of phone-level mispronunciations [7-10]. Among the most common these efforts are pronunciation-scoring based methods [7, 8], of which Goodness of pronunciation (GOP) and its variants are the well-studied representatives. As an illustration, GOP computes the ratio between the likelihoods of the canonical and the most likely pronounced phones which is obtained by aligning the canonical phone sequence of a text prompt with the speech signal uttered by a learner. It then detects phone-level mispronunciations with phone-dependent thresholds. This kind of methods, however, ignore insertion errors in pronunciation segments and fail to give diagnostic feedback. To better obtain informative diagnosis feedback, extended recognition networks (ERN) augments the decoding path of phoneme recognizer with phonological rules [9]. By comparison between an automatic speech recognition (ASR) output and the corresponding text prompt, ERN can readily offer appropriate diagnosis feedback on mispronunciations. Nevertheless, it is practically difficult to enumerate and include sufficient phonological rules into the decoding network for different L1-L2 language pairs.

Recently, end-to-end (E2E) based neural methods have shown promising performance as a competitive alternative for the MDD task [11-13]. A common thought of E2E MDD methods is to first recognize the possible sequence of phones produced by an L2 learner that in turn can be compared to the canonical phone sequence of a text prompt to detect and give feedback on mispronunciations simultaneously. The models of these methods are typically trained to optimize a cross-entropy criterion (viz. maximizing the log-likelihood of the training data), which is not directly linked to the ultimate evaluation metric such as the F1-score employed during the inference phase. Motivated by these observations, we in this paper explore the use of a discriminative objective function for training E2E MDD models, which aims to maximize the expected F1-score directly. As far as we know, our work is the first to explore the training of an E2E MDD model with the maximum F1-score criterion, although there have very limited studies on to optimize the GOP-based models with criteria related to the mispronunciation detection subtask [14, 15]. Extensive experiments on the L2-ARCTIC suggest the utility of our method in relation to some strong baseline MDD models.

The rest of the paper is organized as follows. We first shed light on the instantiation of our E2E MDD model in Section 2. Section 3 elucidates the formulation of maximum F1-score training of an E2E MDD model. After that, the experimental settings and extensive sets of MDD experiments are presented in Sections 4 and 5, respectively. Finally, Section 6 concludes the paper and provides future directions of research.

## 2. HYBRID CTC-ATTENTION MDD MODEL

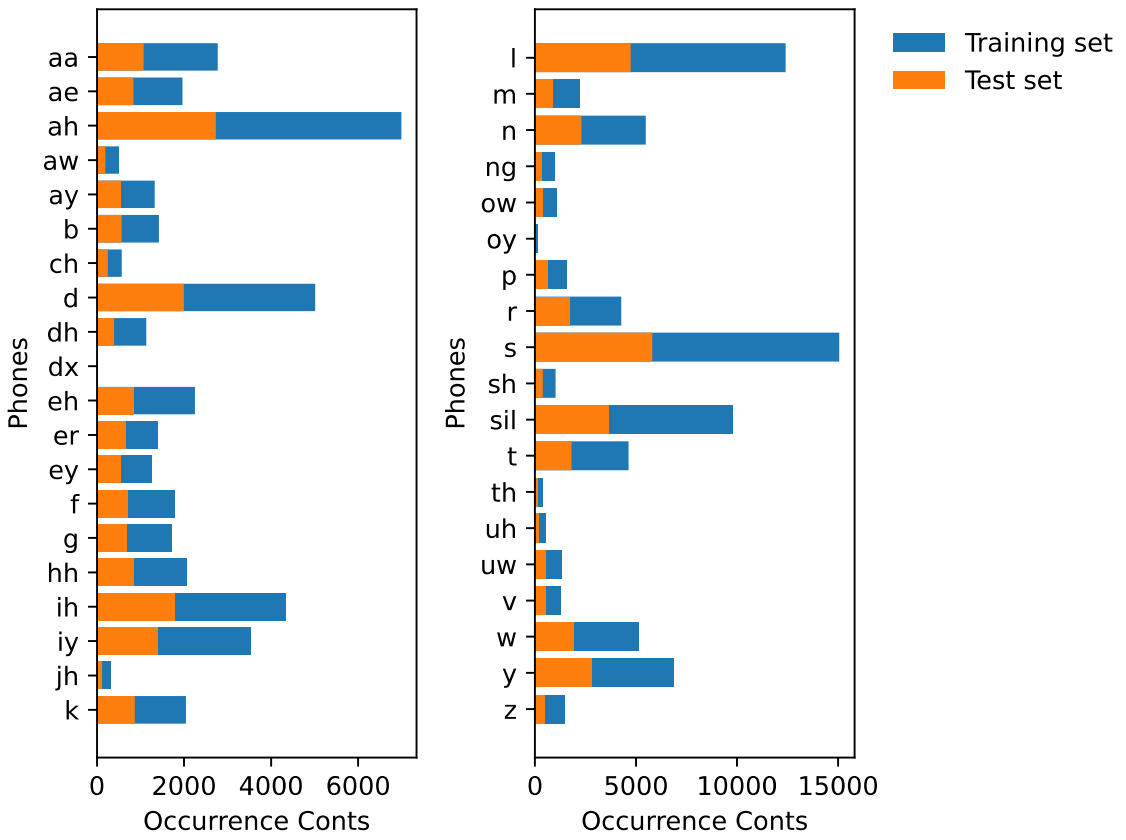

**Fig. 1.** The phonetic distribution of the L2-ARCTIC corpus.

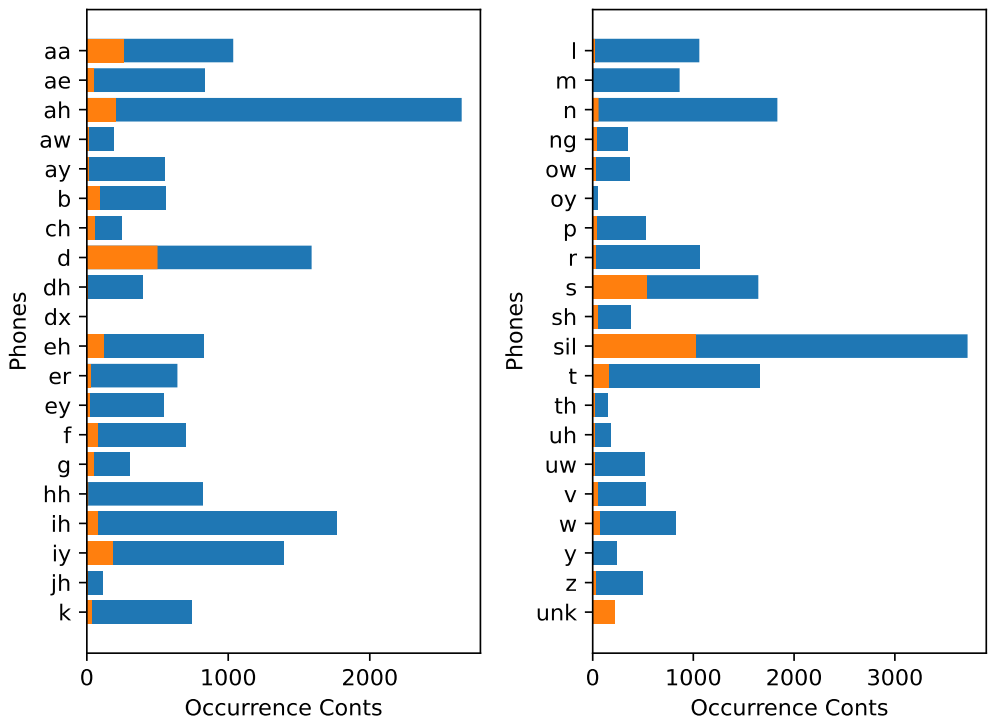

**Fig. 2.** A depiction of the phonetic distribution and pronunciation error patterns of the L2-ARCTIC test set. Each bar shows the occurrence counts of a specific phone and the orange part of each bar indicates the counts that other phones are mispronounced as this phone.

In this section, we describe the E2E model that we capitalize on for MDD, namely the hybrid CTC-Attention model, where CTC is short for Connectionist Temporal Classification [16]. Let $(\mathrm{X}, \mathbf{y}^*)$ denote a training utterance equipped with its annotated transcript of an L1 or L2 speaker, where $\mathrm{X} = \{\mathbf{x}_t \in \mathbb{R}^d \mid t = 1, .., T\}$ is the corresponding acoustic feature vector sequence and $\mathbf{y}^* = \{y_l^* \in \mathcal{V} \mid l = 1, .., L\}$ is the corresponding annotated phone sequence. The training objective of the hybrid CTC-Attention neural model is to maximize a logarithmic linear combination of the posterior probabilities predicted by CTC and the attention-based model, i.e., $P_{CTC}(\mathbf{y}^*|\mathrm{X})$ and $P_{ATT}(\mathbf{y}^*|\mathrm{X})$:

$$LL = \alpha log P_{CTC}(\mathbf{y}^*|\mathrm{X}) + (1-\alpha) log P_{ATT}(\mathbf{y}^*|\mathrm{X}), \quad (1)$$

where $\alpha$ is a linear combination weight. Maximizing Eq. (1) for all training utterances is equivalent to minimize a cross-entropy (CE) loss $\mathcal{L}_{\mathrm{CE}}$. Further, CTC aims to find $\mathbf{y}^*$ that has the maximum likelihood over all valid alignments (each of which is denoted by $A = \{a_t \in \mathcal{V} \cup \{<\text{blank}>\} | t = 1, .., T\}$) between $\mathrm{X}$ and $\mathbf{y}^*$:

$$P_{CTC}(\mathbf{y}^*|\mathrm{X}) = \sum_{A \in \beta^{-1}(\mathbf{y}^*)} P_{CTC}(A|\mathrm{X}), \quad (2)$$

where $\beta^{-1}(\mathbf{y}^*)$ returns all possible alignments compatible with $\mathbf{y}^*$. An alignment $A$ is predicted with the conditional independence assumption between the intermediate output tokens $a_t$:

$$P_{CTC}(A|\mathrm{X}) = \prod_{t=1}^{T} P_{CTC}(a_t|\mathrm{X}). \quad (3)$$

The attention-based model predicts an output token $y_l$ at each time step, conditioning on the previously generated token sequence $y_{1:l-1}^*$ in an autoregressive manner:

$$P_{ATT}(\mathbf{y}^*|\mathrm{X}) = \prod_{l=1}^{L} P_{ATT}(y_l|y_{1:l-1}^*, \mathrm{X}). \quad (4)$$

Note here that for the hybrid CTC-Attention model, CTC works as a regularizer to help eliminate irregular alignments and achieve fast convergence [16].

## 3. MAXIMUM F1-SCORE TRAINING

Discriminative training techniques have been shown to consistently outperform the maximum likelihood paradigm (or the minimum cross-entropy criterion) for acoustic model training in ASR. The most commonly used methods include minimum classification error, maximum mutual information, and minimum phone error [17-19], which generally seek to reducing empirical word (or phone) error rates caused by an ASR system. However, when adapting an ASR modeling framework, in particular the E2E neural framework, to the MDD task, the aforementioned discriminative training criteria do not account well for the ultimate evaluation metrics of mispronunciation detection, such as the F1-score.

In view of this, we explore to further train the model components of our E2E method with a maximum F1-score criterion (MFC), making effective use of the pronunciation error patterns existing in the L2 training utterances. Such a training regime will be better suited for mispronunciation detection and is also expected to boost the performance of the subsequent mispronunciation diagnosis process. The F1-score is a harmonic mean of recall and precision (*cf.* Section 4.2), which can be ultimately expressed below based on the numbers of mispronunciations that are detected by the MDD model, the human assessors, and both of them:

$$\mathrm{F1} = \frac{2 \times C_{D \cap H}}{C_D + C_H}. \quad (5)$$

where $C_D$ is the number of mispronunciations detected by the MDD model, $C_H$ is the number of mispronunciations identified by human assessors and $C_{D \cap H}$ is the number of phone segments marked as mispronunciations by both of the MDD model and the human assessors. Notice that $C_H$ is a constant, while $C_D$ and $C_{D \cap H}$ are computed based on the output of the MDD model.

As such, the training objective for MFC can be defined by

$$\mathcal{L}_{\mathrm{MFC}} = -\frac{1}{n} \sum_{n} \sum_{\mathbf{y}_n^m \in M\mathrm{best}(\mathrm{X}_n)} \mathcal{F}(\mathbf{y}_n^m, \mathbf{y}_{n,}^*, \mathbf{y}_n^+) P_\theta(\mathbf{y}_n^m|\mathrm{X}_n), \quad (6)$$

where for a given L2 training utterance $n$, $\mathbf{y}_n^m$ is a possible phone-level sequence generated by the MDD model $\theta$, $\mathbf{y}_n^*$ is the reference transcription annotated by human assessors and $\mathbf{y}_n^+$ is the canonical phone sequence of the corresponding text prompt; $\mathcal{F}(\mathbf{y}_n^m, \mathbf{y}_{n,}^*, \mathbf{y}_n^+)$ is a function that computes the F1-score of the L2 training utterance $X_n$ given $\mathbf{y}_n^m$, $\mathbf{y}_n^*$ and $\mathbf{y}_n^+$; and $P_\theta(\mathbf{y}_n^m|X_n)$ is approximated by the normalized distribution over *M*-best hypotheses. Accordingly, the MDD model can be optimized to have better F1-score performance by minimizing the loss function defined in Eq. (6) for the L2 training utterances (some of which may contain pronunciation errors).

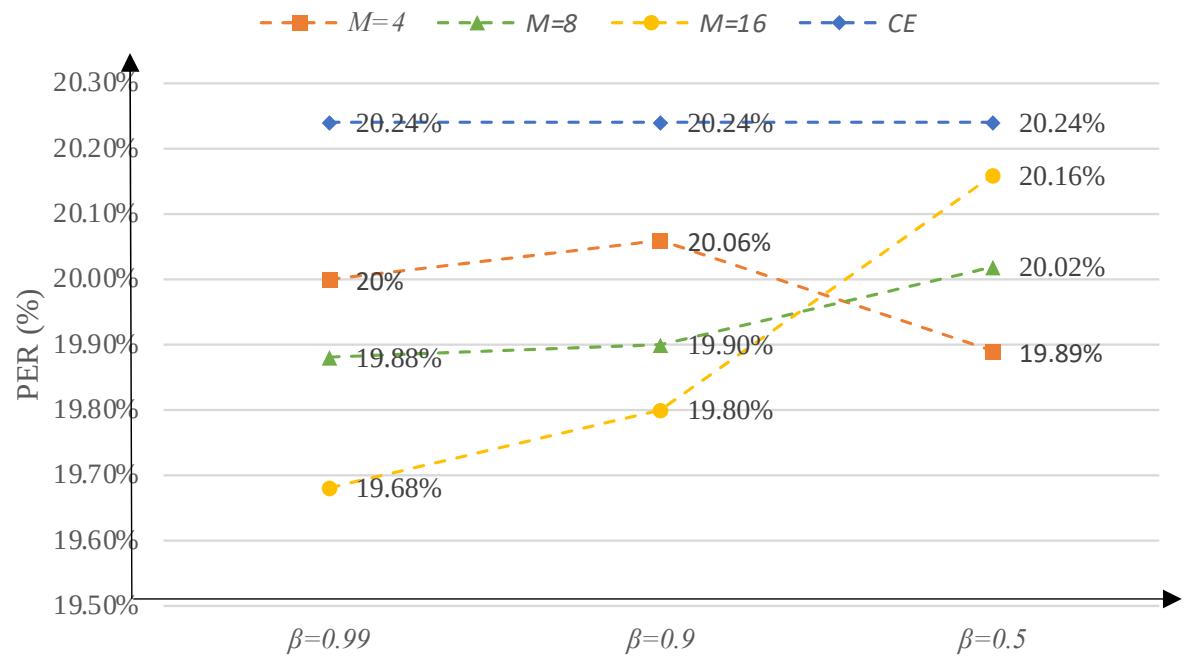

**Fig. 3.** The PER result of our proposed models, as a function of different settings of the *M*-best list and interpolation weight $\beta$.

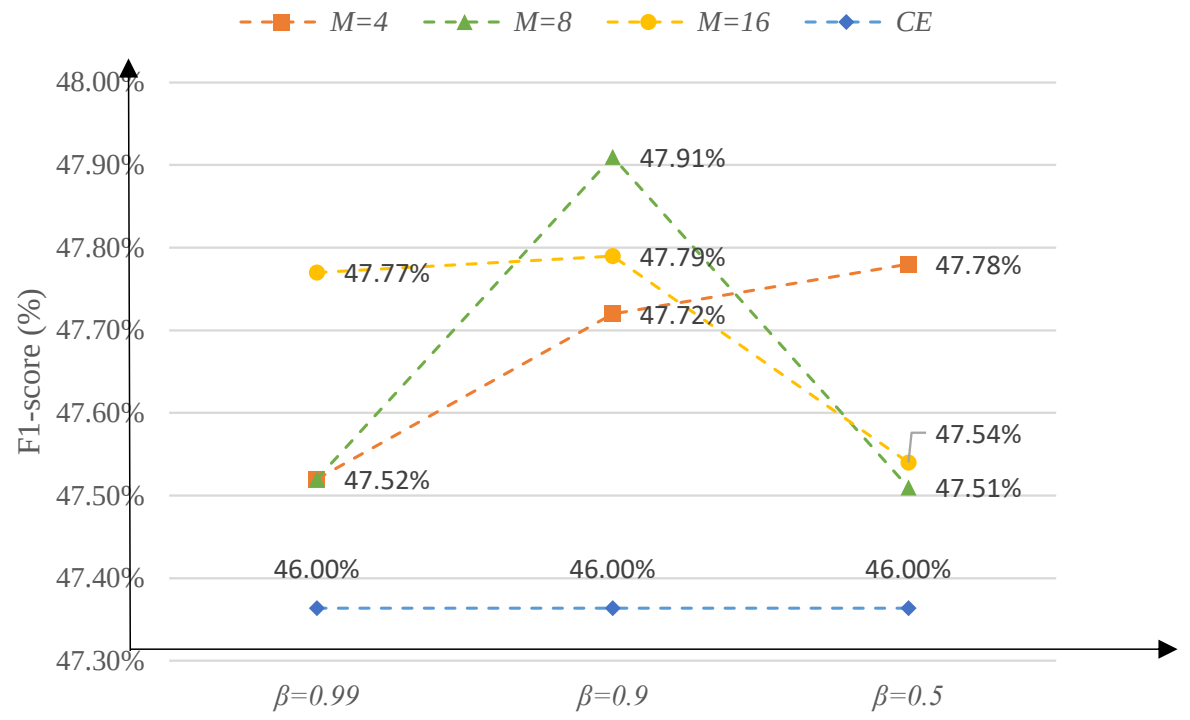

**Fig. 4.** The mispronunciation detection results of our proposed models, as a function of different settings of the *M*-best list and interpolation weight $\beta$.

Finally, the overall loss function $\mathcal{L}$ for training the MDD model is express by

$$\mathcal{L} = \beta \cdot \mathcal{L}_{\text{MFC}} + (1 - \beta) \cdot \mathcal{L}_{\text{CE}}, \quad (7)$$

where $\mathcal{L}_{\text{CE}}$ is the typical cross-entropy loss function operated on the entire L1 and L2 training utterances; and the interpolation weight $\beta$ controls the degree of reliance on $\mathcal{L}_{\text{MFC}}$ rather than $\mathcal{L}_{\text{CE}}$ [20]. The conception of making effective use of evaluation metric-related training criteria has been investigated for the GOP-based method with either a GMM-HMM or DNN-HMM modeling paradigm with some success recently [14, 15]. However, to our knowledge, this conception has yet to be fully explored for E2E MDD methods.

**Table 1.** Statistics of the experimental speech corpora

| | Corpus | Sets | Spks. | Utters. | Hrs. |
|---|---|---|---|---|---|
| L1 | TIMIT | Train | 462 | 3,696 | 3.14 |
| | | Dev. | 50 | 400 | 0.34 |
| | | Test | 24 | 192 | 0.16 |
| L2 | L2-ARCTIC | Train | 15 | 2,249 | 2.29 |
| | | Dev. | 3 | 400 | 0.49 |
| | | Test | 6 | 900 | 0.87 |

**Table 2.** The corresponding confusion matrix for four test conditions of mispronunciation detection

| Total Conditions | | Canonical Pronunciations | |
|---|---|---|---|
| | | CP | MP |
| Model Prediction | CP | True Positive (TP) | False Positive (FP) |
| | MP | False Negative (FN) | True Negative (TN) |

## 4. EXPERIMENTAL SETUP

### 4.1. Speech corpora

We conducted our MDD experiments on the L2-ARCTIC dataset [21]. L2-ARCTIC dataset is an open-access L2-English speech corpus compiled for research on CAPT, accent conversion, and others. L2-ARCTIC contains correctly pronounced utterances and mispronounced utterances of 24 non-native speakers (12 males and 12 females), whose mother-tongue languages include Hindi, Korean, Mandarin, Spanish, Arabic and Vietnamese. A suitable amount of native (L1) English speech data compiled from the TIMIT corpus [22] (composed of 630 speakers) was used to bootstrap the training of the various E2E MDD models and the DNN-HMM acoustic model of the GOP-based method. To unify the phone sets of these two corpora, we followed the definition of the CMU pronunciation dictionary to obtain a set of 39 canonical phones. This phone set was additionally supplemented with a /UNK/ phone corresponding to non-categorial or distortion errors. Next, we divided these two corpora into training, development and test sets, respectively; in particular, the setting of the mispronunciation detection experiments on L2-ARCTIC followed the recipe provided by [12]. Table 1 summarizes the detailed statistics of these speech datasets, while Figure 1 depicts the occurrence counts of each phone in the training and test sets. For the 2,249 utterances of the L2-ARCTIC training set, 11,815 phone substitutions, 113 phone deletions and 113 phone insertions were annotated by the human assessors. As an illustration, the top-5 confusing phone pairs are /z/⟶/s/, /dh/⟶/d/, /ih/⟶/iy/, /d/⟶/sil/ and /er/⟶/ah/ (viz. /z/ was frequently mispronounced as /s/, in term of this analogy for other pairs).

### 4.2. Implementation details

Our baseline E2E MDD method was implemented with the CTC-Attention neural architecture, which was pretrained on the training utterances compiled from the TIMIT corpus and then fine-tuned on the training utterances of the L2-ARCTIC dataset, both with the minimum cross-entropy criterion. On top of this, we further trained the model of our baseline E2E MDD method with the proposed maximum F1-score criterion (*cf.* Section 3) solely on the training utterances of the L2-ARCTIC dataset so as to enhance the discriminative power of our MDD method.

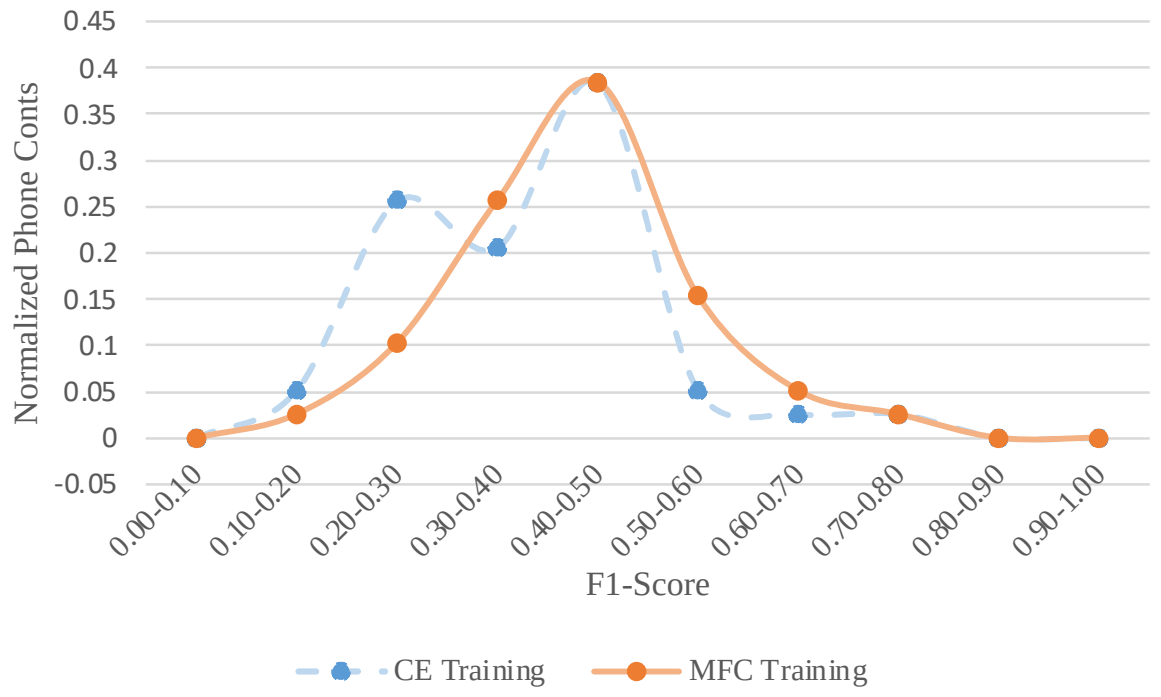

**Fig. 5.** The overall distribution of the F1-scores for performing mispronunciation detection on different kinds of phone (viz., vowel or consonant) segments, based on our baseline model (trained with the CE criterion alone) and best model (additionally trained with the MFC criterion).

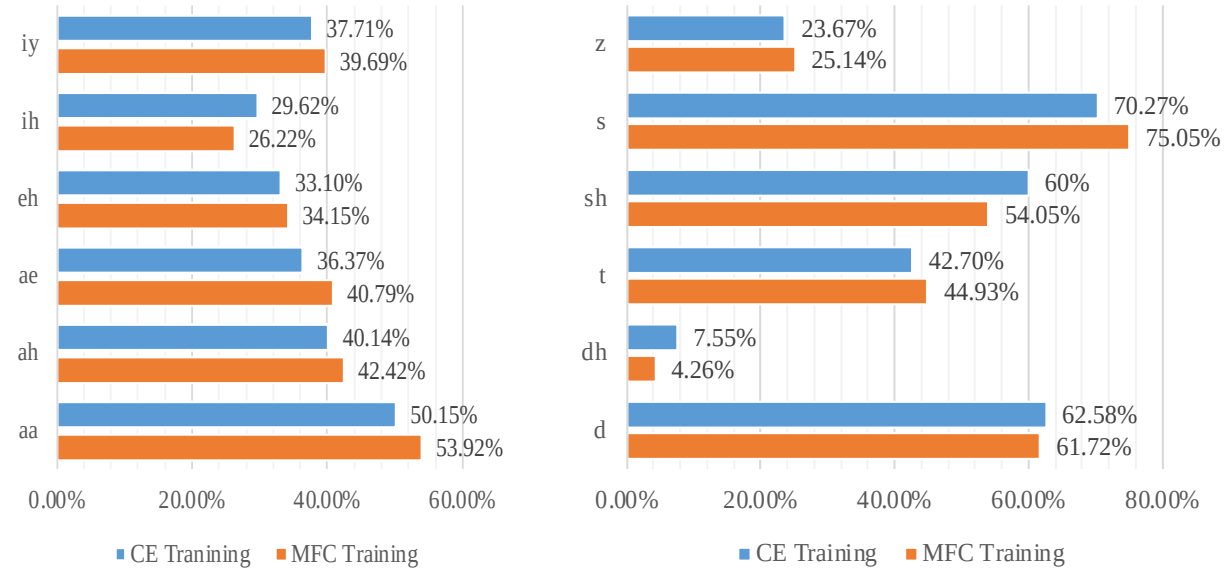

**Fig. 6.** The F1-scores of our proposed training criterion for the most frequent mispronunciation vowels and consonants.

The encoder modules of our various E2E MDD models were all composed of the VGG-based deep CNN (convolutional neural network) component plus a bidirectional LSTM (long short-term memory) component with 1,024 hidden units, while the input to the network is 80-dimensional Mel-filter-bank feature vectors which are extracted every 10 ms with a Hanning window size of 25 ms [16]. Next, the decoder modules of our experimental models all consist of a single-layer LSTM component with 300 hidden units.

### 4.3. Performance evaluation

In order to evaluate the performance of mispronounce detection subtask, we follow the hierarchical evaluation structure adopted in [11]. The corresponding confusion matrix for four test conditions is illustrated in Table 2, where the CP and MP indicate correct pronunciation and mispronunciation conditions, respectively. Based on the statistics (viz. occurrence counts) accumulated from the four test conditions, the metrics for mispronunciation detection can be evaluated with Recall (RE; TN/(FP+TN)), Precision (PR; TN/(FN+TN)) and the F1-score.

**Table 3.** Comparison of the mispronunciation detection and diagnosis performance among different models.

| Models | RE (%) | PR (%) | F1 (%) | DAR (%) |
|---|---|---|---|---|
| GOP | 52.88 | 35.42 | 42.42 | - |
| CNN-RNN-CTC [11] | **76.91** | 32.21 | 45.41 | 42.80 |
| Our Baseline Model | 72.20 | 33.75 | 46.00 | 52.77 |
| Our Best Model | 71.76 | **35.96** | **47.91** | **59.66** |

To calculate the diagnostic accuracy rate (DAR), we focus on the cases of TN and consider it as combination of the numbers of correct diagnosis (CD) and incorrect diagnosis (ID). The accuracy of mispronunciation diagnosis rate (DAR) is calculated by

$$\mathrm{DAR} = \frac{\mathrm{CD}}{\mathrm{CD} + \mathrm{ID}}. \tag{8}$$

## 5. EXPERIMENTAL RESULTS

Before we report on the empirical results of mispronunciation detection and diagnosis experiments for the L2-ARCTIC test set, we perform quantitative analysis on the phonetic distribution and pronunciation error patterns of this test set. By looking at Figure 2, we find that relatively large portions of the occurrence counts of vowels like /aa/, /ah/ and /iy/ are caused due to mispronunciations of other phones as the these vowels, while similar situations exist for consonants like /d/ and /s/.

We then assess the phoneme error rate (PER) of our method. The corresponding results are shown in Figure 3, where the baseline method refers to the hybrid CTC-Attention neural model trained with the minimum cross-entropy (CE) criterion. Considerable PER reductions were achieved by using maximum F1-score criterion (MFC) training. In particular, our method achieves an absolute PER reduction of 0.56% compared to the baseline model when the number of hypotheses in the *M*-best list is set to 16 and the interpolation weight $\beta$ is set to 0.99 for the MFC training (*cf.* Eq. (6)). Further, we proceed to evaluate the task effectiveness of the MFC training as a function of different settings of the *M*-best list and interpolation weight $\beta$, whose F1-score results are depicted in Figure 4. It is evident that the MFC training can further boost the mispronunciation detection performance of the MDD model trained with the CE criterion by a significant margin; especially, the setting of *M*=8 and $\beta = 0.9$ for the MFC training yields the highest gain. Here we look at the other side of the coin, viz. the overall distribution of the F1-score results for mispronunciation detection of different kinds of phone segments, based either on our baseline model (trained with the CE criterion alone) or on our best MDD model (additionally trained with the MFC criterion). The corresponding distributions are visualized in Figure 5. We can find that there is a marked shift of the distribution to the direction of higher F1-scores when our model was further trained with the MFC criterion, apart from the CE criterion. Going one step further, we compare the F1-scores of some most frequently mispronounced vowels and consonants for these two models. As shown in Figure 6, our proposed MFC training criterion can consistently improve the F1-scores of the most frequently mispronounced vowels (except for the /ih/); however, for the most frequently mispronounced consonants, a mixed view is presented, which deserves further investigation.

In the last set of experiments, we first compare the results of our best model to that of the well-practiced GOP-based method [7] and the recently proposed E2E MDD models on mispronunciation detection. Inspection of Table 3, we can find, on the whole, that the E2E models can yield better F1-score results than the GOP-based method, while our best model outperforms CNN-RNN-CTC in terms of the F1-score metric. Next, we turn to studying the efficacy of our best model on the mispronunciation diagnosis subtask, in relation to our baseline model (viz. CTC-ATT trained with the CE criterion) and CNN-RNN-CTC. From the last column of Table 3, we can also confirm the merits of the MFC training for improving the discriminative capability of the E2E MDD model for mispronunciation diagnosis.

## 6. CONCLUSION AND FUTURE WORK

In this paper, we have designed and developed a novel maximum F1-score criterion (MFC) for training E2E mispronunciation detection and diagnosis (MDD) models, taking the hybrid CTC-Attention model as an example. An extensive set of MDD experiments conducted on the L2-ARCTIC dataset seems to confirm the utility of our methods. An interesting direction for future work is to explore more sophisticated E2E neural models and effective training criteria for MDD. Furthermore, we also plan to extend our model to make use of suprasegmental-level information cues for CAPT, such as intonation, pitch and rhythm.

## 7. ACKNOWLEDGEMENTS

This research is supported in part by Chunghwa Telecom Laboratories under Grant Number TL-110-D301, and by the National Science Council, Taiwan under Grant Number MOST 109-2634-F-008-006- through Pervasive Artificial Intelligence Research (PAIR) Labs, Taiwan, and Grant Numbers MOST 108-2221-E-003-005-MY3 and MOST 109-2221-E-003-020-MY3. Any findings and implications in the paper do not necessarily reflect those of the sponsors.